\documentclass[floats,floatfix,amssymb,prd,letter,aps,nofootinbib,nolongbibliography,reprint,superscriptaddress]{revtex4-2}

\usepackage{ulem} 
\usepackage{changes}
\usepackage{amssymb,amsmath,verbatim,mathtools,needspace,enumitem,etoolbox,graphicx,physics,microtype,afterpage,xspace,tabularx,lmodern,multirow}
\usepackage{gensymb}
\usepackage[dvipsnames, usenames]{xcolor}
\usepackage[unicode, colorlinks=true, linkcolor=linkcolor, citecolor=linkcolor, filecolor=linkcolor, urlcolor=linkcolor, linktocpage, breaklinks]{hyperref}
\newcommand{\orcid}[1]{\href{https://orcid.org/#1}{\includegraphics[width=10pt]{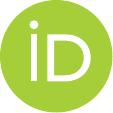}}}
\usepackage{orcidlink}
\usepackage[nolist,nohyperlinks]{acronym}
\usepackage{tikz}
\usetikzlibrary{decorations.markings}
\usetikzlibrary{decorations.pathmorphing}

\definecolor{linkcolor}{rgb}{0.6, 0, 0.5} \definecolor{olive}{rgb}{0.4,0.7,0.2}
\newif\ifshowcomments

\showcommentsfalse 

\newif\ifcommenthiddenprinted
\commenthiddenprintedfalse

\newcommand{\commenthidden}{%
  \ifcommenthiddenprinted
  \else
    \global\commenthiddenprintedtrue
  \fi
}

\newcommand{\mda}[1]{\ifshowcomments\textcolor{purple}{\textbf{MD}: #1}\else\commenthidden\fi}
\newcommand{\gc}[1]{\ifshowcomments\textcolor{orange}{\textbf{GC}: #1}\else\commenthidden\fi}

\newcommand{\ls}[1]{\ifshowcomments{\color{teal}{{\bf LS}: #1}}\else\commenthidden\fi}

\newcommand{\newsection}[1]{\textit{\textbf{#1.}}}

\newcommand{\centra}{CENTRA, Departamento de Fisica, Instituto Superior T\'ecnico – IST,
Universidade de Lisboa – UL, Avenida Rovisco Pais 1, 1049 Lisboa, Portugal}

\newlength{\bigfigwidth}
\allowdisplaybreaks

\begin{document}

\pagenumbering{arabic}


\title{Dynamical quasinormal mode excitation II:\\ propagation and convergence in Schwarzschild}
\author{Marina De Amicis~\orcidlink{0000-0003-0808-3026}}
\email{mdeamicis@perimeterinstitute.ca}
\affiliation{Perimeter Institute for Theoretical Physics, 31 Caroline St N, Waterloo, ON N2L 2Y5}
\author{Enrico Cannizzaro~\orcidlink{0000-0002-9109-0675}}
\affiliation{\centra}
\author{Gregorio Carullo~\orcidlink{0000-0001-9090-1862}}
\affiliation{School of Physics and Astronomy and Institute for Gravitational Wave Astronomy, University of Birmingham, Edgbaston, Birmingham, B15 2TT, United Kingdom}
\author{Adrien Kuntz~\orcidlink{0000-0002-4803-2998}}
\affiliation{\centra}
\author{Laura Sberna~\orcidlink{0000-0002-8751-9889}}
\affiliation{School of Mathematical Sciences, University of Nottingham, University Park, Nottingham NG7 2RD, United Kingdom}

\begin{abstract}
We study the dynamical excitation of quasinormal modes (QNMs) during the plunge of a particle into a Schwarzschild black hole, building on the framework of \textit{Phys. Rev. D 113 (2026) 2, 024048} (Paper I).
Investigating the high-frequency behavior of Leaver's QNM solutions, we obtain a more accurate and general prescription for their propagation.
We confirm the existence of a new ``characteristic radius'' for QNM excitation, the \textit{bounce radius} $r_*=0$, in agreement with recent literature. 
To its right, the QNM signal scatters off this point before reaching the observer; to its left, it propagates directly on the light-cone. 
Applying the formalism of Paper I to inspiralling particles, and 
using this refined prescription, we obtain a QNM signal that accurately reproduces the oscillatory component of the waveform after the bounce crossing, yielding an essentially complete first-principles description of the waveform from shortly after the signal peak. 
The dynamical QNM signal undergoes a transition as the particle crosses the bounce radius: from a quasi-resonant regime, where successive overtones are driven in counter-phase and interfere destructively, to a free-oscillator one, where they are in phase and the QNM sum converges rapidly.
These results provide a clear physical interpretation of the collective QNM behavior during the plunge, and a firm theoretical foundation for accurate ringdown modelling.

\end{abstract}

\maketitle

\newsection{Introduction}
Standard ringdown templates employed in black hole (BH) spectroscopy~\cite{Berti:2025hly} are based on ``one-body'' BH perturbation theory~\cite{Chandrasekhar:1975zza,Leaver:1985ax,Leaver:1986gd,PhysRevD.55.468,Berti:2009kk}, describing the post-merger gravitational waveform as a superposition of damped sinusoids with constant amplitude coefficients, the quasinormal modes (QNMs). 
The constant-amplitude assumption is strictly valid only for narrow Gaussian initial data on a fixed BH background; for more general initial data, mode amplitudes are not constant even within the one-body setting, once propagation effects are properly accounted for~\cite{Chavda:2024awq}.
More broadly, this ``stationary'' picture has a limited domain of validity in realistic scenarios: numerical analyses~\cite{Kamaretsos:2012bs,London:2018gaq,Baibhav:2023clw,Mitman:2025hgy,Berti:2025hly} show that it holds only some time after the waveform peak, while extending it to earlier times requires including a large number of modes, at the risk of overfitting. 
To address this theoretical limitation, standard merger-ringdown templates are expressed through phenomenological ans\"atze, numerically calibrated in terms of the binary degrees of freedom, both in the extreme mass-ratio limit~\cite{Albanesi:2023bgi,Faggioli:2026alx, Albanesi:2026qtx} and for comparable-masses~\cite{Damour:2014sva, Bohe:2016gbl, Nagar:2018zoe, Cotesta:2018fcv, Nagar:2019wds, Nagar:2020pcj,Estelles:2020osj, Estelles:2020twz, Gamba:2021ydi, Pompili:2023tna}. 
An interesting avenue is offered by the self-force program which, while not providing a closed-form expression for the ringdown waveform, can provide a perturbative numerical mapping between the plunge properties and the stationary ringdown amplitudes \cite{Hughes:2019zmt,Lim:2019xrb,Becker:2025zzw}, and waveforms valid throughout the evolution of small-mass-ratio systems~\cite{Kuchler:2025hwx}.
Paper I~\cite{DeAmicis:2025xuh} modelled the ringdown of a two-body system in the small-mass-ratio limit, retaining one-body perturbation theory tools while accounting for a plunging particle source.
It investigated how the QNM signal propagates with respect to the curved light-cone, leveraging a Green’s function
separation into a prompt response (and tail) piece plus a QNM (and tail) contribution valid for generic source locations, inspired by the large-distance approximation of Ref.~\cite{PhysRevD.55.468}.
According to Paper I, QNMs travel on hyperboloidal slices in the minimal gauge, a prescription smoothly interpolating between light-cone propagation and scattering off $r_*=0$ for sources localized, respectively, close and far from the BH horizon.
Leveraging this propagation prescription (therein denoted as ``QNM causality condition''), Paper I studied how QNMs are dynamically excited when a particle plunges into a Schwarzschild BH.
Among other results, it showed for the first time that: 
i) for a realistic source, the amplitudes are time-dependent throughout the plunge and until well after the light-ring crossing; 
ii) before the light-ring crossing, QNM excitation behaves as an oscillator driven by the particle’s orbital frequency; 
iii) at late times, the QNM signal emitted by a source approaching the horizon is redshifted, leading to the presence of new \textit{redshift terms}, non-oscillating and exponentially damped with decay rates given by multiples of the horizon surface gravity.\footnote{The redshift terms of Paper I are a different spectral feature than the ``horizon modes'' of Refs.~\cite{Mino:2008at,Zimmerman:2011dx,Oshita:2025qmn}, see Ref.~\cite{Rosato:2026moe}.} In a QNM-only template, redshift terms would appear as overtones amplitudes growing at late times.
Subsequent work studied the Green's function for the Pöschl-Teller potential~\cite{Kuntz:2025gdq,Arnaudo:2025uos} and Schwarzschild-de Sitter spacetime~\cite{Arnaudo:2025uos}, identifying all its components beyond the QNM sector. While Ref.~\cite{Kuntz:2025gdq} qualitatively reproduced the source-driven QNM excitation of Paper I, both works adopted a different prescription for QNM propagation -- one that switches sharply between scattering and light-cone propagation at a certain \textit{bounce radius} (coinciding with the potential peak for P$\rm \ddot{o}$schl-Teller), and is identical with Paper I's only when the source is close to or far from the horizon. Ref.~\cite{Arnaudo:2025uos} further argued that its Green's function reduces to the Schwarzschild one in the limit of vanishing cosmological constant $\Lambda$, and Ref.~\cite{Arnaudo:2025kit} showed, by explicitly taking $\Lambda\to 0$, that the QNM propagation condition of Refs.~\cite{Kuntz:2025gdq,Arnaudo:2025uos} (with bounce radius now at $r_*=0$) is necessary for the convergence of the Price's law component. The same prescription was employed in Ref.~\cite{Su:2026fvj}, which numerically studied the spectral decomposition of the Schwarzschild Green's function for a source at large distance from the BH, building on a high-frequency analysis of Leaver's QNMs~\cite{Casals:2011aa}.
Given this recent literature, we revisit the assumptions underlying the QNM propagation prescription used in Paper I, and show that one of them was too restrictive. 
Its relaxation leads to a new condition, in agreement with Refs.~\cite{Casals:2011aa,Kuntz:2025gdq,Arnaudo:2025uos,Arnaudo:2025kit}: a sharp transition from scattering off $r_*=0$ to propagation inside the lightcone.
We then go beyond recent results, using the framework of Paper I to recompute the particle-driven dynamical excitation of QNMs with this updated prescription.
Our previous results remain qualitatively valid, but we can now fully reproduce the oscillatory component of the numerical signal after the particle crosses the bounce radius.
Additionally, we obtain a new picture of collective modes excitation, showing that each mode is excited with 
close to opposite (equal) phase for subsequent overtones, before (after) the bounce crossing.

\newsection{Black hole perturbations}
\label{sec:framework}
We consider linearized perturbations on top of a Schwarzschild background with horizon radius $r_H$.
We expand the gravitational wave strain in spin-2 weighted spherical harmonics ${}_{-2}Y_{\ell m}(\Theta,\Phi)$, with $(\Theta,\Phi)$ angular coordinates of the observer.
We construct two gauge-invariant master functions: the Zerilli, $\Psi^e_{\ell m}$, and the Regge-Wheeler, $\Psi^o_{\ell m}$, variables, encoding respectively the even
and the odd 
sectors under parity transformations.
These functions satisfy wave-like equations of the form
\begin{equation}
    \left[\partial_t^2-\partial_{r_*}^2+V^{e/o}_{\ell}(r)\right]\Psi^{e/o}_{\ell m}(t,r_*)=S^{e/o}_{\ell m}(t,r_*) \, ,
    \label{eq:RWZ_sourced}
\end{equation}

where $r_* = r + r_H \log (r/r_H-1)$ is the tortoise coordinate,  $V^{e/o}_{\ell }$ is the potential barrier of the background while $S^{e/o}_{\ell m}$ is the source generating the perturbations. 
In this work, we will consider perturbations driven by an infalling particle, whose source can be found in Paper I and Ref.~\cite{Nagar:2006xv}. We initialize the particle sufficiently far away in time from the merger, and far from the BH, so that the initial data can be neglected.

\newsection{Numerical code}
To evolve the trajectory of the particle all the way to the plunge-merger, we use the numerical code \textsc{RWZHyp}~\cite{Bernuzzi:2010ty,Bernuzzi:2011aj}, using the same settings as Paper I.
This code solves for the Hamiltonian equations of motion of the particle, driven by radiation-reaction effective forces, see Ref.~\cite{DeAmicis:2024not} for more details.
The code also computes the numerical waveform emitted by the system, solving the Regge-Wheeler/Zerilli sourced equations in Eq.~\eqref{eq:RWZ_sourced}.
Numerical waveforms are computed at future null infinity $\mathcal{I}^+$ through a hyperboloidal layer parametrized by retarded time $\tau$ and a compactified radial coordinate $\rho$, related to the particle coordinates $(t,r_*)$ through $\tau-\rho_+=t-r_*$, with $\rho_+$ location of $\mathcal{I}^+$. 
We work in geometric units $c=G=1$ and we rescale dimensional quantities with respect to $M=r_H/2$.

\newsection{Green's function approach}
The general solution of Eq.~\eqref{eq:RWZ_sourced} can be found through convolution with the retarded Green's function associated to the Regge-Wheeler/Zerilli differential operator
\begin{equation}
\Psi_{\ell m}^{(e/o)}(t,r_*)=\int_{-\infty}^{\infty}dt'\int_{-\infty}^{\infty}dr_*' \ G_{\ell m}(t-t';r_*,r_*')S_{\ell m}(t',r_*') \, .
\end{equation}
The Green's function $G$ is the solution of Eq.~\eqref{eq:RWZ_sourced} with impulsive source $S^{\rm imp}_{\ell m}=\delta(t-t')\delta(r_*-r_*')$ subject to dissipative boundary conditions, i.e. purely ingoing/outgoing plane waves at the horizon/spatial infinity.
The time evolution of the Schwarzschild Green's function is characterized by three distinct features: an early-time response~\cite{Kuntz:2025gdq,Arnaudo:2025uos,DeAmicis:2026tus,Su:2026fvj,PhysRevD.55.468} that can be approximated as a polynomial in the observer's retarded time~\cite{DeAmicis:2026tus}; an intermediate-time ringing, given by the superposition of the BH's QNMs~\cite{Leaver:1985ax}; a late-time inverse power law decay, usually referred to as ``tail''~\cite{Price:1971fb,Leaver:1986gd}.
Each of these time-domain features is associated to different spectral components of the Green's function in the frequency domain.
The Laplace-transformed Green's function\footnote{Our Laplace transform  convention is $\tilde{\phi}(\omega)=\int_{t_0}^{\infty} \mathrm{d} t \, e^{i\omega t}\phi(t)$, with $t_0$ the initial time.} 
is written as
\begin{align}
\tilde{G}_{\ell m}(\omega;r_*,r_*')=&\frac{-1}{W_{\ell m}(\omega)} \left[\theta(r_*-r_*')u_{\ell m}^{\rm in}(\omega,r_*')u_{\ell m}^{\rm up}(\omega,r_*)\right. \nonumber \\
& + \left.   \theta(r_*'-r_*)u_{\ell m}^{\rm in}(\omega,r_*)u_{\ell m}^{\rm up}(\omega,r_*')  \right] ,
\label{eq:GF_freq}
\end{align}
where $u_{\ell m}^{\rm up, in}$ are two independent solutions of the $\omega$-domain homogeneous Regge-Wheeler/Zerilli equation and $W_{\ell m}$ is their Wronskian.
In the following, we will suppress the $\ell m$ indices to avoid clutter.
In particular, $u^{\rm in}$ behaves as unitary, ingoing plane wave $\sim e^{-i\omega r_*}$ at the horizon, and as superposition $\sim A_{\rm in}(\omega) e^{-i\omega r_*} + A_{\rm up}(\omega) e^{i\omega r_*} $ at large distances. 
Instead, $u^{\rm up}$ is a purely outgoing plane wave at spatial infinity $u^{\rm up}\sim e^{i\omega r_*}$, and a combination of ingoing and outgoing plane waves at the horizon. 
With this notation, the Wronskian is $W = 2 i \omega A_\mathrm{in}$.
It is useful to introduce a third solution $u^{\rm down}$ to the homogeneous, $\omega$-domain Regge-Wheeler/Zerilli problem, defined as purely ingoing at large distances $u^{\rm down}\sim e^{-i\omega r_*} \, , r_*\rightarrow\infty$, and combination of both ingoing and outgoing modes at the horizon. 
The three solutions are not all linearly independent: $u^{\rm in}$ can be rewritten as combination of the up- and down-modes $u^{\rm in}(\omega,r_*)=A_{\rm in}(\omega) u^{\rm down}(\omega,r_*)+A_{\rm up}(\omega) u^{\rm up}(\omega,r_*)$.
This allows to rewrite the Green's function~\eqref{eq:GF_freq} as a function only of $u^{\rm up/down}$.
In particular, assuming that the observer is at $\mathcal{I}^+$ and further from the BH than the source ($r_*>r_*'$), we can split this quantity in two components $\tilde{G}(\omega;r_*,r_*')=\tilde{G}^{(1)}(\omega;r_*,r_*')+\tilde{G}^{(2)}(\omega;r_*,r_*')$, defined as
\begin{equation}
    \begin{split}
        &\tilde{G}^{(1)}(\omega;r_*,r_*')\equiv \frac{i}{2\omega}e^{i\omega r_*}u^{\rm down}(\omega,r_*') \, , \\
        &\tilde{G}^{(2)}(\omega;r_*,r_*')\equiv \frac{i}{2\omega}\frac{A_{\rm up}(\omega)}{A_{\rm in}(\omega)}e^{i\omega r_*}u^{\rm up}(\omega,r_*')\, .
    \end{split}
\end{equation}

The quasinormal frequencies (QNFs) correspond to zeros of $A_{\rm in}$ in the complex $\omega$-plane, hence poles of the Green's function component $G^{(2)}$.
The time-domain contribution of $G^{(2)}$ to the full retarded propagator is
\begin{equation}
\begin{split}
    G^{(2)}&(t-t';r_*,r_*')=\\& \frac{i}{2}\int_{-\infty+i\epsilon}^{+\infty+i\epsilon}\frac{\mathrm{d} \omega}{2 \pi} \, e^{-i\omega(t-r_*-t')} \, \frac{A_{\rm up}(\omega)}{\omega A_{\rm in}(\omega)}u^{\rm up}(\omega,r_*')\, ,
\end{split}
\label{eq:G2_integral}
\end{equation}
with $\epsilon\in \mathbb{R}^+$. 
The QNFs are limited to the lower half of the complex $\omega$-plane, and appear in the time-domain signal only when it is possible to close the complex contour in this region, a step necessary to compute Eq.~\eqref{eq:G2_integral} using the residue theorem.
As proposed in Paper I (see its Section~III), to understand whether the contour in Eq.~\eqref{eq:G2_integral} can be closed in the lower half plane, it is necessary to study the behavior of the integrand for high overtones $\omega_n, \, n\gg1$.
Studying this limit, Paper I argued that the QNM response in $G^{(2)}$ is, in general, activated with a time delay with respect to $G^{(1)}$.
It denoted this time delay ``QNM causality condition'', proposing a closed-form expression later challenged in Refs.~\cite{Kuntz:2025gdq,Arnaudo:2025uos,Su:2026fvj}. 
Here, we show that the ``causality condition'' of Paper I is indeed too conservative: one of the hypotheses behind its derivation can be relaxed, leading to a condition, in agreement with Refs.~\cite{Casals:2011aa,Kuntz:2025gdq,Arnaudo:2025uos,Su:2026fvj}.
Hereafter, we refer to the time at which the QNM contribution in $G^{(2)}$ is activated as the \textit{bounce time}, and reserve the term \textit{causality condition} for the arrival time of the first signal propagated by the retarded Green's function, i.e. the lightcone condition $t-r_*=t'-r_*'$.
This notation, consistent with previous literature~\cite{Arnaudo:2025uos}, is supported by a physical intuition, as discussed in the next section.
\newsection{Quasinormal modes high-frequency limit}
\label{sec:high_freq}
\begin{figure}[t]
\centering
\includegraphics[trim={0cm 0.65cm 0cm 0.cm},width=0.49\textwidth]{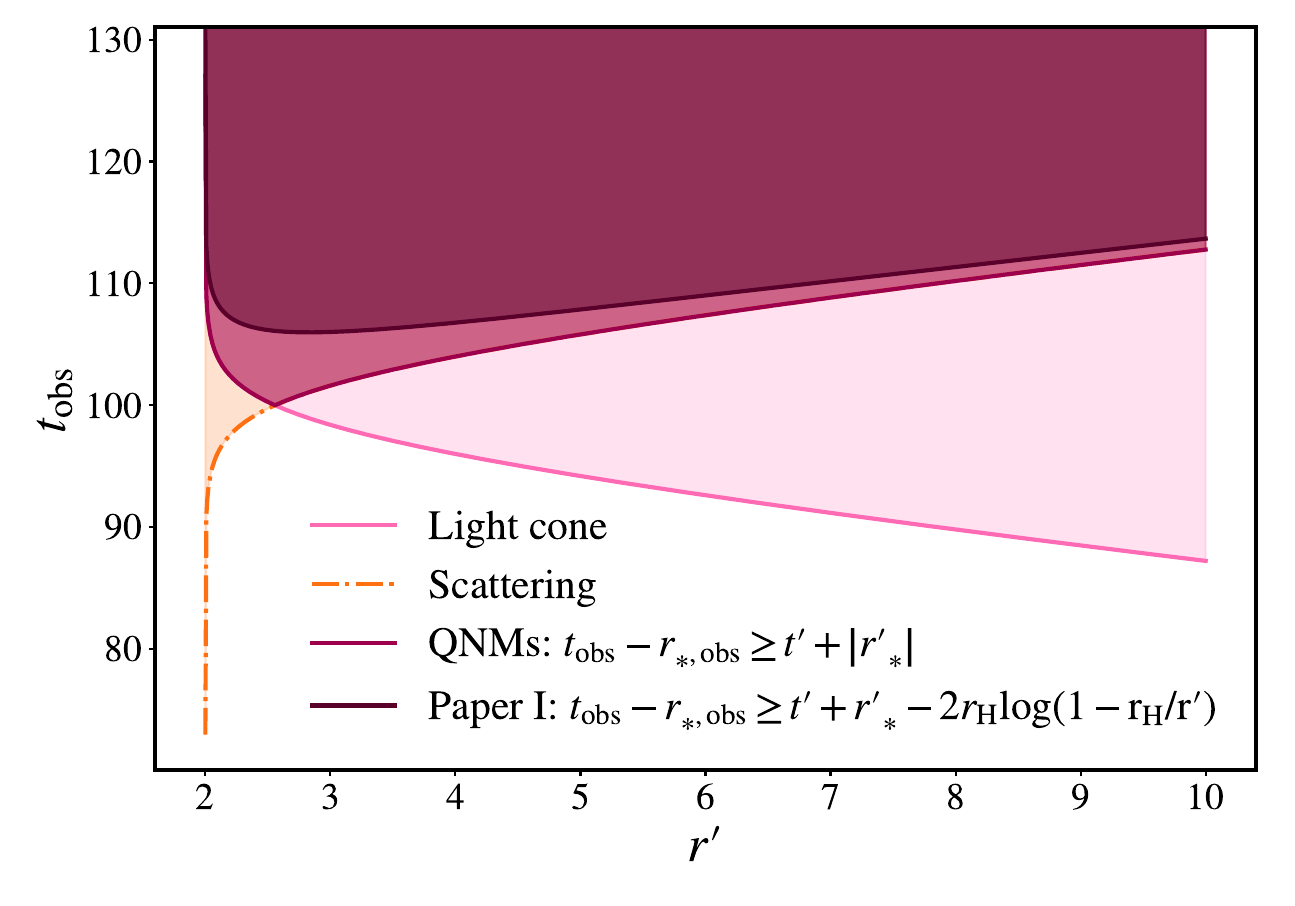}
\caption{
Time $t_{\rm obs}$ at which an observer at $r_{\rm obs}=100M$ receives a signal (assuming $t'=0$): traveling on the curved lightcone $t_{\rm obs}-r_{*,\rm obs}=t'-r_*'$ (pink); scattering from near the potential barrier peak $t_{\rm obs}-r_{*,\rm obs}=t'+r_*'$ (orange);  travelling on the portion of the curved lightcone over which the QNMs propagate (light purple). We compare this prescription with Paper I's (dark purple).
}
\label{fig:lightcone}
\end{figure}
Near the QNFs $\omega\approx\omega_n$, the argument of Eq.~\eqref{eq:G2_integral} behaves as
\begin{equation}
\begin{split}
    &e^{-i\omega(t-r_*-t')}\tilde{G}^{(2)}(\omega;r_*,r_*')
    \bigg|_{\omega\approx\omega_n}\simeq\\ 
    &\qquad \frac{i B_n}{\omega-\omega_n} \frac{\hat{a}(\omega_n,r_*')}{4\pi \omega_n} e^{-i\omega(t-r_*-t'-\mathcal{C}(r_*'))}\, ,
\end{split}
\end{equation}
where
\begin{align}
& B_n\equiv A_{\rm up}(\omega_n)\times\left[\frac{dA_{\rm in}(\omega)}{d\omega}\bigg|_{\omega=\omega_n}\right]^{-1} \, , \nonumber \\
& \hat{a}_n(r')\equiv \left[\sum_ka_k(\omega_n)\left(1-\frac{r_H}{r'}\right)^k\right]\times\left[\sum_k a_k(\omega_n)\right]^{-1} \, , \nonumber \\
&  \mathcal{C}(r_*')\equiv r_*'-2r_H\log\left(1-\frac{r_H}{r'}\right)  \, ,
\label{eq:BnHataCold_defs}
\end{align}
and with $a_k$ the near-horizon-expansion coefficients of Leaver's solution for the quasinormal eigenfunctions~\cite{Leaver:1985ax}. 
In investigating the behavior of this integrand in the large overtone number $n$ limit, Paper I made the following assumptions:
\begin{equation}
\begin{split}
 & \lim_{n\rightarrow \infty}\frac{\log|B_n|}{\mathrm{Im}(\omega_n)}=0  \, , \\
 &  \lim_{n\rightarrow \infty}\frac{\log|\hat{a}_n(r')|}{\mathrm{Im}(\omega_n)}=0  \ \, , \ \forall \, r'  \, .
\end{split}
\label{eq:assumptions}
\end{equation}
Under these hypotheses, the requirement that the integrand in Eq.~\eqref{eq:G2_integral} is well behaved for $\mathrm{Im}(\omega)\rightarrow-\infty$ reduces to asking that
\begin{equation}
  e^{-i\omega_n(t-r_*-t'-\mathcal{C}(r_*'))} \xrightarrow[n\to\infty]{} 0 \, ,
\end{equation}
which yields the QNM propagation condition used in Paper I,
\begin{equation}
    t-r_*\geq t'+\mathcal{C}(r_*') \, .
\end{equation}
The first assumption in Eq.~\eqref{eq:assumptions} is guaranteed by the behavior of the QNFs, $\omega_n\propto - i \, n$ \cite{Motl:2002hd}, and excitation factors $B_n\sim \mathcal{O}(1)$ at large $n$, see App.~A of Ref.~\cite{Berti:2006wq} (accounting for our different definition of the latter).
The second hypothesis in Eq.~\eqref{eq:assumptions} is valid for compact sources far from the BH, since in this case $\hat{a}(r'\gg 1)\rightarrow 1$ and $\hat{a}(r')$ does not contain terms scaling as $\sim e^{i\omega r'}$.
Based on this limit, Paper I assumed the condition to hold for arbitrary source locations.
However, through an improved \textsc{Mathematica} implementation of Leaver's algorithm~\cite{Leaver:1985ax, Nollert:1993zz}, now accurate up to $n=70$ overtones, we find that this is not the case.
In App~\ref{app:high-freq}, we show numerically that 
\begin{equation}
|\hat{a}_n(r')| \xrightarrow[n\to\infty]{} |e^{-i\omega_n\, (\mathcal{C}(r_*')-|r_*'|)}|\, ,
\label{eq:Delta_def}
\end{equation}
\gc{the appendix, and the equation below, has $\mathrm{Im}(\omega_n)$, not $|\mathrm{Im}(\omega_n)|$.}
with $\mathcal{C}(r_*')$ introduced in Eq.~\eqref{eq:BnHataCold_defs}. 
Then, the integrand in Eq.~\eqref{eq:G2_integral} is well behaved for high frequencies $\mathrm{Im}(\omega)\rightarrow-\infty$ if the following condition holds 
\begin{equation}
  \hat{a}_n(r')e^{-i\omega_n(t-r_*-t'-\mathcal{C}(r_*'))}\sim e^{-i\omega_n(t-r_*-t'-|r_*'|)}\rightarrow 0 \, ,
\end{equation}
yielding a different prescription for when the signal propagated by QNMs reaches an observer at $(t,r_*)$, 
\begin{equation}
    t-r_*\geq t'+|r_*'| \, ,  \, \, \forall r_*'\, ,
    \label{eq:new_caus}
\end{equation}
in agreement with Refs.~\cite{Berti:2006wq,Casals:2011aa,Kuntz:2025gdq,Arnaudo:2025uos,Su:2026fvj}. In App.~\ref{app:high-freq}, we provide strong hints that the QNM series converges inside the curved light-cone portion selected by Eq.~\eqref{eq:new_caus}.
We compare different propagation conditions in Fig.~\ref{fig:lightcone}.
Note that, since the propagation condition \eqref{eq:new_caus} depends on the asymptotics of the mode functions at large $n$, it is independent of the harmonic number $\ell$.
Following this prescription, QNMs emitted by a compact source localized in the $r_*'>0$ region can reach $\mathcal{I}^+$ only after scattering off $r_*'=0$, which for this reason we refer to as the \textit{bounce radius}.
If the compact source is localized in $r_*'<0$, Eq.~\eqref{eq:new_caus} becomes the causality condition of propagation on/inside the light-cone.
Note also that (contrary to the P$\mathrm{\ddot{o}}$schl-Teller case~\cite{Kuntz:2025gdq}) $r_*=0$ ($r \simeq 2.55$) does not coincide with the peak of the potential in Eq.~\eqref{eq:RWZ_sourced}, nor with the light ring. It constitutes a new ``characteristic radius'' in the physics of QNM excitation.

\newsection{Particle-driven quasinormal modes}
\label{sec:result}
\begin{figure}[t]
\centering
\includegraphics[trim={0cm 1.05cm 0cm 0.3cm},width=0.48\textwidth]{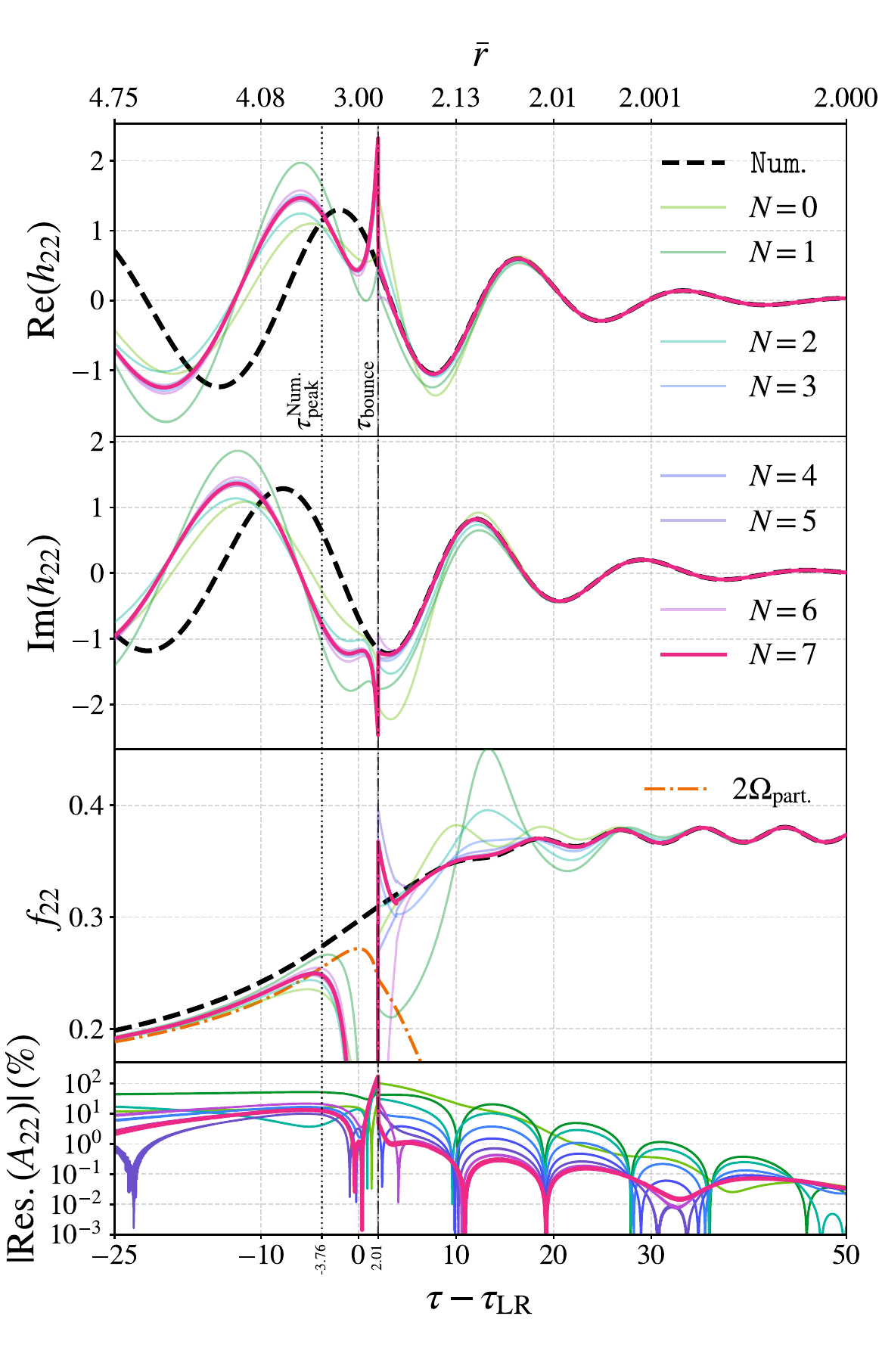}
\caption{
Waveform (first and second row) and instantaneous frequency (third row) as a function of the retarded time $\tau$ of the observer at $\mathcal{I}^+$ measured from light-ring crossing time $\tau_{\rm LR}$, for a quasi circular inspiral.
We show the QNM signal obtained adding up $N$ of overtones and their mirror modes (colors) and the numerical solution (dashed black).
Theoretical predictions have been aligned in phase to the numerical results, while $\Omega_{\rm part}$ is the particle orbital frequency.
The last row shows the residuals between the numerical and the QNM-propagated waveform amplitudes, defined as $|\mathrm{Res.}(A_{22})|=100 \cdot |A_{22}^{\rm num}-A_{22}^{\rm QNMs}|/A_{22}^{\rm num}$.
The vertical lines represent the times $\tau^{\rm Num}_{\rm peak}$ at which the numerical waveform amplitude peaks (black dotted) and $\tau_{\rm bounce}$ at which $\bar{r}_*(\tau)=0$ (dot-dashed).}
\label{fig:wf_ecc0}
\end{figure}
We now study the dynamical ringdown regime, following the formalism introduced in Paper I, but with the prescription \eqref{eq:new_caus} for QNM propagation.
While the expression of the activation coefficient is the same as in Paper I,

\begin{align}
    c_n(\tau)=&e^{i\omega_n \rho_+}\int_{\bar{r}_*}^{\infty}dr_*'\, |\dot{r}_*(t(r_*'))|^{-1}\,\lbrace u_n(t',r')\, f(t',r') \nonumber \\
    &\left.-\left(1-r_H/r'\right)\,\partial_{r'}\left[g(t',r')u_n(t',r')\right]\right\rbrace_{t'=t(r_*')} ,
\end{align}
the impulsive coefficient contains a derivative of the QNM bounce time, so it becomes
\begin{equation}
  i_{n}(\tau)= e^{i\omega_n \rho_+}\frac{u_n(t(\bar{r}),\bar{r})\, g(t(\bar{r}),\bar{r}) \, \mathrm{sign}(\bar{r}_*)}{|1+\mathrm{sign}(\bar{r}_*) \, \dot{r}_*(\bar{r})|} \, .
\end{equation}
with the convention $\rm sign(0)=0$.
In the above expressions, we have introduced the apparent location of the source $\bar{r}$ as the solution of
\begin{equation}
\tau-\rho_+-t(\bar{r})-|r_*(\bar{r})|=0\, .
\label{eq:causality_trajectory}
\end{equation}
where $\tau$ is the retarded time.
We focus on the $\ell=m=2$ multipole of a plunging quasi-circular inspiral ($e_0=0.0$, Table~I of Paper I).
Other configurations with intermediate eccentricities yield similar results to those obtained for the quasi-circular case, and are omitted for brevity. For completeness, we discuss a radial infall in App.~\ref{app:rad_infall}.
Since the activation/impulsive coefficients and their full signal contribution are similar to the results of Paper I, we discuss them in App.~\ref{app:coeffs}.
In Fig.~\ref{fig:wf_ecc0} we compare the numerical perturbative waveform with the signal propagated by the QNM Green's function.
We show the waveform, instantaneous frequency and residuals between the predicted and numerical amplitudes. 
The predictions are obtained by adding impulsive $\zeta_{n\pm}$ and excitation $\psi_{n\pm}$ contributions (see Eqs.~\eqref{eq:psi_def} and~\eqref{eq:zeta_def}), with increasing number of overtones.
\begin{figure}[t]
\centering
\includegraphics[width=0.5\textwidth, trim={0cm 1.cm 0cm 0.7cm}]{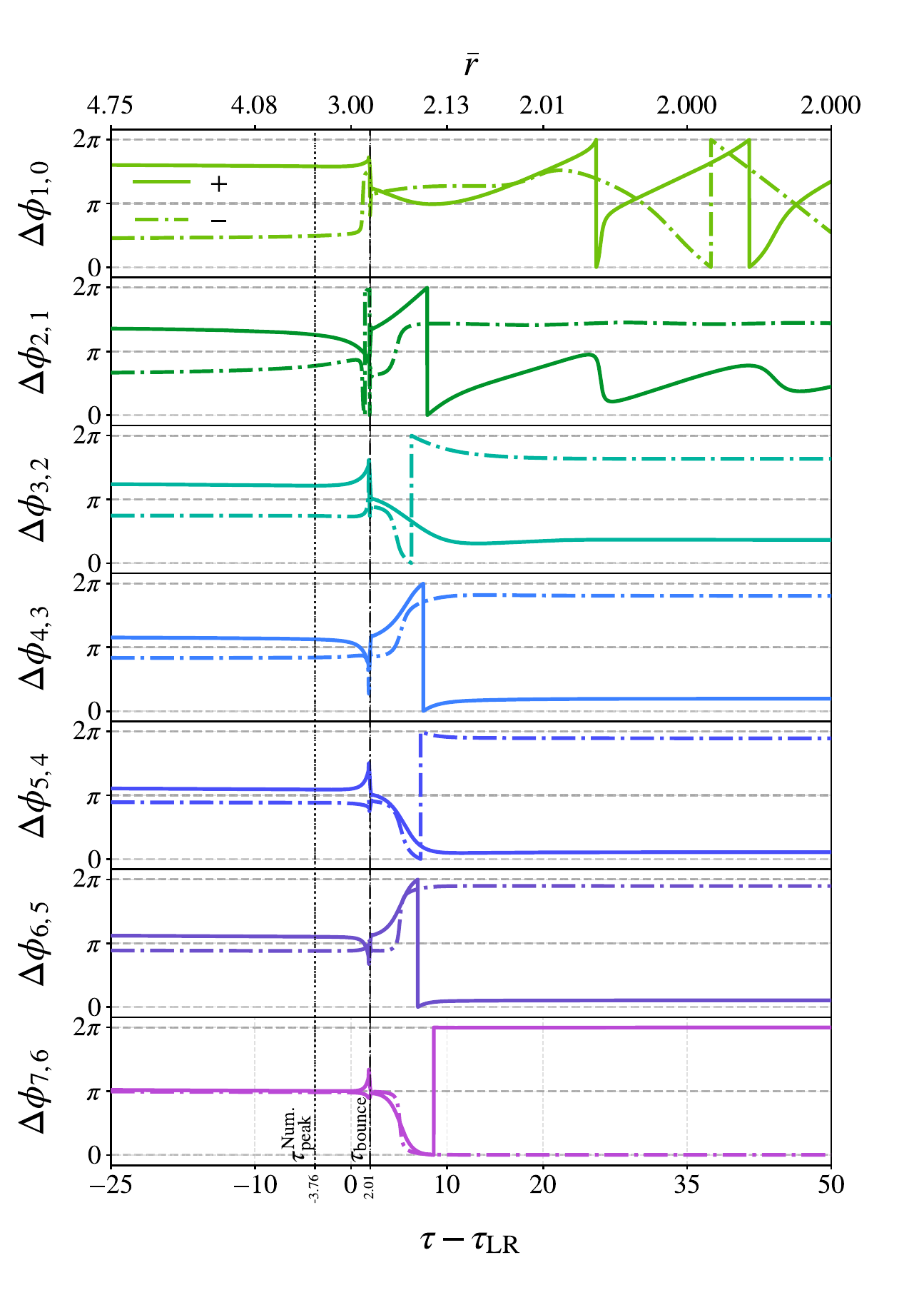}
\caption{Relative phase of successive overtone contributions to the waveform (as defined in Eq.~\eqref{eq:DeltaPhi_def}, and assuming [0,$2\pi$] wrapping), as a function of the retarded time, for a quasi circular inspiral.
}
\label{fig:overt_phases}
\end{figure}
The agreement between the predicted and numerical waveforms strongly depends on the apparent location of the particle with respect to the crossing $\bar{r}_*=0$. 
At that point, the QNM propagation suddenly switches from scattering (off the bounce radius $\bar{r}_*=0$), to propagation inside the lightcone. 
Before this location is crossed, the QNM-propagated signal alone cannot reproduce the full numerical waveform, as other components (namely, the prompt response) are also active. 
This is because for times such that $\bar{r}_*(\tau)>0$, 
the contributions of $G^{(2)}$ are delayed with respect to the ones of $G^{(1)}$, with the delay occupied by the prompt response.
In the total Green's function, the prompt has a large magnitude and is expected to greatly contribute to the full signal, i.e. the convolution of $G^{(1)}$ and $G^{(2)}$ with the particle source (see also Ref.~\cite{Ma:2026qbq}). 
After the particle crosses $\bar{r}_*=0$, $G^{(1)}$ and $G^{(2)}$ are activated simultaneously: the prompt-response term in $G^{(1)}$ is exactly canceled by the one in $G^{(2)}$ at all times, and the only surviving contributions to the full Green's function are the QNM and late-time tail component of $G^{(2)}$~\cite{Kuntz:2025gdq,Arnaudo:2025uos,Su:2026fvj,Ma:2026qbq}. 
Indeed, right after $\bar{r}_*=0$, corresponding to $\tau \simeq \tau_{\rm LR}+2$ for the quasi-circular case, the dynamical QNM waveform matches the numerical one.
A clarification is in order: for retarded times $\tau$ right after crossing $\bar{r}_*=0$, while the full signal is accurately reproduced by the QNM signal alone, it is still highly dynamical and the amplitudes of each mode are not yet constant, see App~\ref{app:coeffs}.
In the quasi-circular case under study, $N > 7$ is necessary for the QNM signal to converge during the plunge and before $\bar{r}_*(\tau)=0$; convergence sharply improves after this point.
This is visible in Fig.~\ref{fig:wf_ecc0}: the amplitude residuals are large and oscillatory in the number of overtones for $\bar{r}_*(\tau)>0$, while for $\bar{r}_*(\tau)<0$ they are suppressed as more and more overtones are added. 
Higher overtones become progressively less significant after the particle crosses $\bar{r}_*=0$. However, note that at intermediate times $\tau-\tau_{\rm LR}\in[2,30]$, the first six overtones are all significantly excited; only after this time the residuals become progressively insensitive to the addition of more overtones.
Studying the instantaneous frequency in Fig.~\ref{fig:wf_ecc0}, we see that the intuition behind the QNM excitation developed in Paper I also holds with the corrected prescription of Eq.~\eqref{eq:new_caus}, with minor differences related to the role of $\bar{r}_*=0$.
Indeed, before the bounce-radius crossing the collective behavior of the QNMs is driven by the source, which oscillates at the particle orbital frequency. The latter grows during plunge, peaks at light-ring crossing, and rapidly decays after. 
This behavior is mimicked by the QNM signal's instantaneous frequency until the time at which the waveform peaks; after, this frequency rapidly decays until $\bar{r}_*=0$.
Soon after, the behavior changes: the QNM signal's instantaneous frequency rapidly settles onto the fundamental mode frequency, and the QNM signal behaves as a free oscillator.

\newsection{Dynamics of relative mode phases}
In Fig.~\ref{fig:overt_phases}, we show the relative phases of the contribution of each overtone to the full waveform
\begin{align}
\Delta\phi_{n+1, n}\equiv & \, \mathrm{Arg}\left\lbrace[c_{n+1}(\tau)+i_{n+1}(\tau)]e^{-i\omega_{n+1} (\tau-\rho_+)}\right\rbrace \nonumber \\ &-\mathrm{Arg}\left\lbrace[c_{n}(\tau)+i_{n}(\tau)]e^{-i\omega_{n} (\tau-\rho_+)}\right\rbrace \,,
\label{eq:DeltaPhi_def}
\end{align}
for the quasi-circular trajectory.
Before the light ring crossing, relative phases are approximately constant and approach $\sim\pi$ as the overtone number increases. 
That is, there is destructive interference between successive overtone signals -- see also the real and imaginary part of the signal generated by each QNM in Fig.~\ref{fig:overt_phases_re_im}, in App~\ref{app:single_mode}.
Long enough after the light-ring crossing, for $n>1$, the relative phases are approximately constant and grow closer to $\sim0$ (modulo $2\pi$) with $n$.
An increasing/decreasing alternated pattern is also present at late times for $n>2$, while $n=1$ presents a non-trivial oscillating behavior.
A subtlety in the interpretation of Fig.~\ref{fig:overt_phases} is that at late times the redshift terms contaminate the QNM-propagated waveform. 

\newsection{Discussion}\label{sec:discussion}
The results we obtain for the QNM excitation and impulsive coefficients of each mode and the corresponding waveform are broadly similar to those of Paper~I. 
Key differences are:
i) The excitation and impulsive coefficient, together with resulting QNM-propagated signal, are not smooth at the bounce-crossing time; 
ii) Before this time, we do not observe convergence with increasing overtone number (up to $n=7$); 
iii) After the bounce time, the prompt response vanishes and the overall agreement between our dynamical QNM prediction and the numerical waveform improves drastically: the waveform is accurately described by dynamical QNMs, with a tail contribution only prominent in radial infalls (as expected~\cite{DeAmicis:2024not}).
In this work, we found another interesting feature, following from a sharp transition between collective and independent mode excitation around the bounce crossing. 
Before (after) the bounce-crossing, subsequent overtones are approximately excited in counter-phase (with aligned-phases), with relative phases approaching $~\sim \pi$ ($\sim 0$) for increasing overtone numbers.
There have been claims that high overtones (up to $n \sim7$) are excited after the waveform peak with opposite phases~\cite{Giesler:2019uxc,Giesler:2024hcr,Coleman:2025ipc}, leading to a destructive interference that suppresses their contribution to the full signal at this stage. 
These studies employed numerical fits of comparable-mass quasi-circular binary mergers waveforms, with a large number of modes. 
Fig.~\ref{fig:overt_phases} potentially explains these results: in a small time interval after the waveform peak the relative phases of each overtone appear in counter-phase. However, this happens in an interval during which the QNM amplitudes and phases show a time dependence.
Further work is needed for a more accurate comparison between first-principles results in the small-mass-ratio limit and fits to nonlinear simulations.
Aside from dynamical QNM excitation, Paper I introduced an additional component, the ``redshift terms'', and studied the dependence of the stationary, late-time amplitude of the fundamental mode in terms of orbital eccentricity.
These two results are not influenced by the change in prescription for QNM propagation at intermediate radii, since they are computed at late enough time (in the near horizon regime), when both prescriptions reduce to the light cone (see Fig.~\ref{fig:lightcone}).
The results presented in this work open the way for many future directions of investigations, which were comprehensively discussed in Sec.~VIII of Paper I.
Here, we note that our new results bring an essentially complete and first-principles waveform description past the bounce crossing, soon after the signal peak.
This brings a significant improvement: we showed there is a time, the bounce-crossing time, after which only QNMs (and possibly the tail) make up the waveform, albeit with time-dependent amplitudes.
Once our results are extended to spinning BHs, we expect this fact to bring significant clarity on ongoing debates on the physics and data analysis of QNMs, and in particular overtone contributions.

\textit{\textbf{Note:}} 
While this work was in preparation, Ref.~\cite{Ma:2026qbq} numerically computed the signal emitted by a toy-model source of a particle infalling into a Schwarzschild BH, using the same prescription for QNM propagation as in Refs.~\cite{Casals:2011aa,Kuntz:2025gdq,Arnaudo:2025uos} and the present work, and a numerical expression for the prompt response.
Its focus was to show that prompt response and QNMs together can completely reproduce a signal computed through direct numerical integration, independently obtaining results qualitatively similar to ours in the QNM component. 
%

\textit{\textbf{Acknowledgements -- }} 
We acknowledge instrumental discussions with Paolo Arnaudo, Lorenzo K\"uchler, Adam Pound and Benjamin Withers on the QNM causality condition and bounce radius. 
M. De Amicis would like to thank L. Lehner for useful conversations.
L.~Sberna would also like to thank S.~Green and J.~Lestingi for helpful discussions. 
A.K. thanks the Fundação para a Ciência e Tecnologia (FCT), Portugal, for the financial support to the Center for Astrophysics and Gravitation (CENTRA/IST/ULisboa) through grant No. UID/PRR/00099/2025 and grant No. UID/00099/2025, as well as to the FCT project ``Gravitational waves as a new probe of fundamental physics and astrophysics'' grant agreement 2023.07357.CEECIND/CP2830/CT0003.
E.C. acknowledges financial support
provided under the European Union’s H2020 ERC Advanced Grant “Black holes: gravitational engines of discovery” grant agreement no. Gravitas–101052587, and from the
Villum Investigator program supported by the VILLUM
Foundation (grant no. VIL37766) and the DNRF Chair
program (grant no. DNRF162) by the Danish National
Research Foundation. L.S.~is supported by a University of Nottingham Anne McLaren Fellowship. 
This research was supported in part by
Perimeter Institute for Theoretical Physics. Research at Perimeter Institute is supported in part by the Government of Canada through the Department of Innovation, Science and Economic Development and by the Province of Ontario through the Ministry of Colleges and Universities.
This work is supported by the Simons Foundation International \cite{sfi} and the Simons Foundation \cite{sf} through Simons Foundation grant SFI-MPS-BH-00012593-11.


\clearpage

\appendix

\onecolumngrid

\section{High frequency limit of $\hat{a}_n(r')$ and convergence of the QNM series}\label{app:high-freq}
\begin{figure}[t]
\centering
\includegraphics[width=0.49\columnwidth]{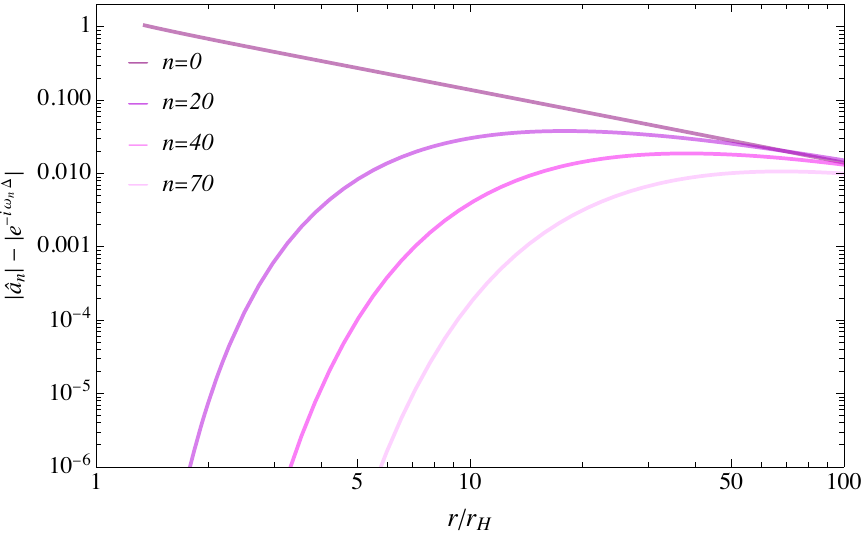}
\includegraphics[width=0.47\columnwidth]{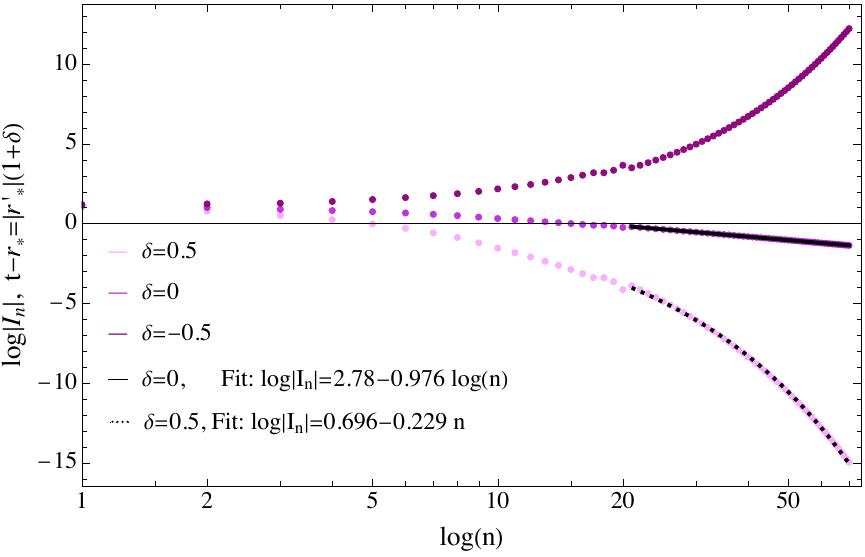}
\caption{\ls{something wrong with the legend in the left panel?}\mda{what should i change? i do not see the error now}
\textbf{Left:} Difference between $\hat{a}_n(r)$, Eq.~\eqref{eq:BnHataCold_defs}, and the exponential behavior in Eq.~\eqref{eq:Delta_def}, versus the horizon-rescaled radius. 
Different colors correspond to values computed for different QNM frequencies, labelled by the overtone number $n$.
\textbf{Right:} Trend vs $\log(n)$ of the $n$-element $I_n=\hat{a}_n(r')e^{-i\omega_n(t-r_*-t'-\mathcal{C}(r_*'))}$  of the series in Eq.~\eqref{eq:series_In}, computed at three different values of the observer's retarded time, as defined in the caption and legend. \ls{These are discrete values, maybe better to use points rather than a line?}\mda{it looks really bad, especially when compared with the fits...}
}
\label{fig:hatavsN_and_ConvvsLogN}
\end{figure}
In Fig.~\ref{fig:hatavsN_and_ConvvsLogN} (left), we study the behavior of Leaver's near-horizon series $\hat{a}(r')$ as defined in Eq.~\eqref{eq:BnHataCold_defs}, as a function of $r'$, for different overtones $n=0,20,40,70$.
We compute the expansion coefficients using two independent implementations of Leaver’s continued-fraction method~\cite{Leaver:1985ax}, one of which also employs Nollert's improvement for the asymptotic recurrence relation~\cite{Nollert:1993zz}; the two approaches agree to machine precision.
We show that, when progressively increasing the overtone number, $\hat{a}(r')$ approaches the following exponential behavior in $r'$
\begin{equation}
e^{-i\omega_n \Delta(r_*')}\equiv e^{-i\omega_n(\mathcal{C}(r_*')-|r_*'|)}  \, ,
\label{eq:Delta_def}
\end{equation}
yielding a pure plane wave behavior that switches sharply between in/out-going for $r_*<0$/$r_*>0$, modulo the term $e^{-i\omega_n\mathcal{C}(r_*')}$, i.e., the low-overtone exponential behavior of Leaver's QNMs.
It is possible to obtain a strong numerical hint that the series below 
\begin{equation}
    \sum_n I_n\equiv\sum_n \hat{a}_n(r')e^{-i\omega_n(t-r_*-t'-\mathcal{C}(r_*'))} \, ,
    \label{eq:series_In}
\end{equation}
converges when $t-r_*>t'+|r_*'|$. 
To show this, we plot in Fig.~\ref{fig:hatavsN_and_ConvvsLogN} (right) the quantity $\log|I_n|$ as a function of $n$, with the source at $(t'=0,r_*'=3M)$, and at three different retarded times of the observer $t-r_*=|r_*'|(1+\delta)$.
For $\delta=0$, we test the convergence at the time the first QNM signal is supposed to arrive according to the revised propagation condition of Eq.~\eqref{eq:new_caus}, while for $\delta=\pm 0.5$ we consider slightly later/earlier times.
As we consider an observer retarded time $t-r_*$ inside the prescription in Eq.~\eqref{eq:new_caus} (i.e., $\delta=0.5$), we find that there exists an overtone number $\bar{n}$, such that $\log|I_{n>\bar{n}}|$ is well fitted by a linear behavior in $n$ with negative slope.
This hints, as far as we can say with a limited number of overtones $n\leq 70$, that for $n>\bar{n}$, subsequent terms in the series $\sum_n I_n$ decay exponentially, which is a sufficient condition for the series to converge.

For $t-r_*$ outside the prescription in Eq.~\eqref{eq:new_caus} (i.e., $\delta=-0.5$), there exists a value $\bar{n}$ after which subsequent terms in the series $\sum_nI_n$ start growing in $n$, which suggests that the series does not converge.
The case $t-r_*=|r_*'|$ (i.e., $\delta=0$) is marginal: for $n\leq70$, it appears that the series in Eq.~\eqref{eq:series_In} is not convergent. In fact, for all the $n$ we were able to test, numerical fits show that $\log|I_n|\simeq b-\alpha \log n$ with $0<\alpha\lesssim 1$.  
The series we are analyzing in Eq.~\eqref{eq:series_In} is not the one appearing in the time-domain Green's function, since it lacks the factor $B_n/\omega_n$
\begin{equation}
    G_{\rm QNM}(t-t';r_*,r_*')\simeq
    \sum_n\frac{B_n}{2\omega_n}\hat{a}(\omega_n,r_*') e^{-i\omega(t-r_*-t'-\mathcal{C}(r_*'))}\, .
    \label{eq:QNMs_series}
\end{equation}
Since for high overtone numbers $\omega_n\propto -(i/4) \, (n+1/2)$~\cite{PhysRevD.55.468}, the factor $\omega_n^{-1}$ alone does not impact the exponential convergence for $t-r_*>t'+|r_*'|$. However, it might affect the convergence of $\sum_n I_n \omega_n^{-1}$ at times $t-r_*=t'+|r_*'|$. The same reasoning can be applied to the factor $B_n$, since it does not involve an exponential behavior in $n$~\cite{Berti:2006wq}.

\clearpage

\section{QNM-waveform driven by a radial infall}\label{app:rad_infall}

\begin{figure}[t]
\centering
\includegraphics[trim={0.6cm 0.65cm 0cm 0.cm},width=0.45\textwidth]{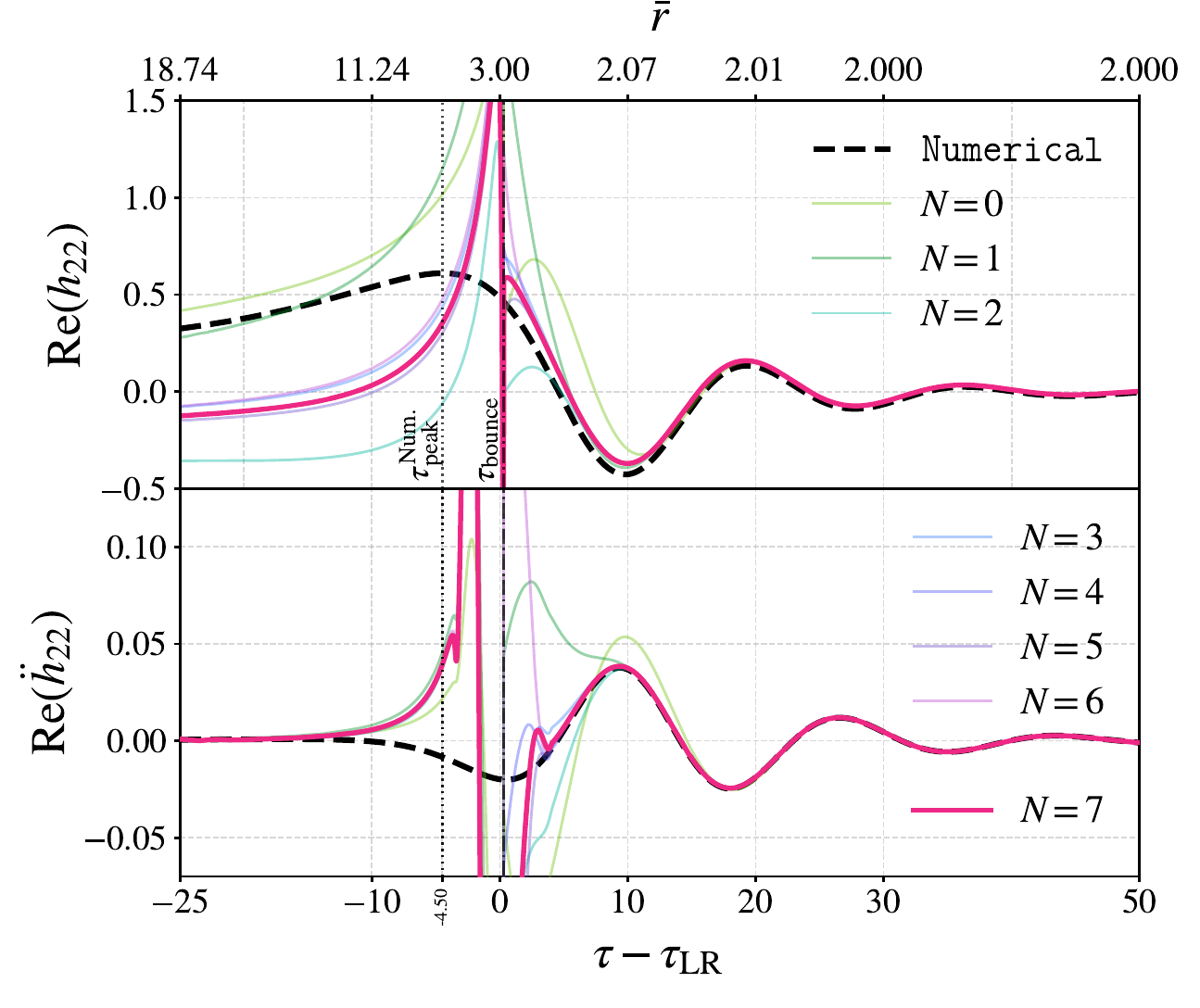}
\includegraphics[trim={0.6cm 0.65cm 0cm 0.cm},width=0.45\textwidth]{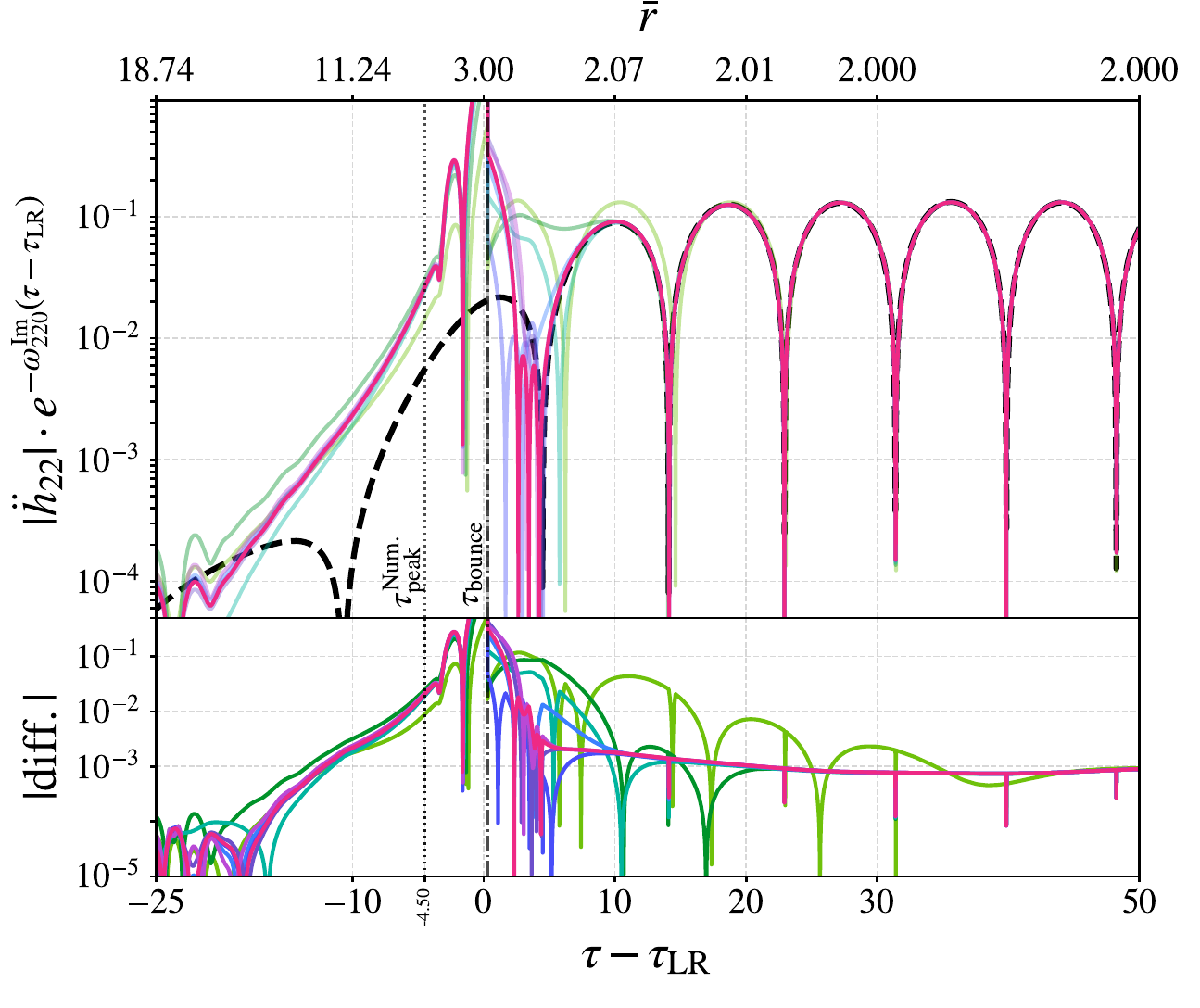}
\caption{
\textbf{Left}:
Strain quadrupole $h_{22}$ (top) and its second derivative $\ddot{h}_{22}$ (bottom) vs retarded time of the observer at $\mathcal{I}^+$, measured from the light-ring crossing time. \textbf{Right}: Rescaled quadrupole second derivative (top) and difference between its numerical value and the one propagated by QNMs, vs the retarded time.
Results relative to a \textbf{radial infall} from $r_0=50$ with initial particle energy (rescaled by its mass) $E_0=1$. 
Colors as in Fig.~\ref{fig:wf_ecc0}.
The vertical black dotted/dot-dashed line represents the retarded time of the quadrupole peak $\tau_{\rm peak}$/$\tau_{\rm bounce}$ at which $\bar{r}_*(\tau)=0$.
}
\label{fig:wf_radinf}
\end{figure}
We study the quadrupolar ($\ell=m=2$) waveform emitted by a particle infalling from $r_0=50$, with initial energy $E_0=1$, as rescaled with its mass.
Activation and impulsive coefficients and contributions relative to this case are shown in App~\ref{app:coeffs}.
In Fig.~\ref{fig:wf_radinf}, we compare the only non-vanishing polarization of the QNM-propagated signal vs the numerical waveform, and the residuals of their amplitudes.
A close agreement after $\bar{r}_*=0$ ($\tau=\tau_{\rm LR}+0.287$ in this case) is present only in $\ddot{h}$, while a mismatch appears when analyzing the strain.
This mismatch is due to the tail term, suppressed close to the merger for a quasi circular plunge, but highly enhanced by a radial infall~\cite{DeAmicis:2024not,Rosato:2026moe}.
The time derivatives suppress power-laws with respect to exponentials, hence the better agreement in $\ddot{h}$.
We argue that the mismatch close to $\bar{r}_*=0$ in the $\ddot{h}_{22}$ signal, is in part an effect of the numerical second derivative itself, since the predicted QNM signal is not smooth at this crossing.
We see from the residuals in Fig.~\ref{fig:wf_radinf} that fewer overtones are excited by a radial infall, compared with the quasi-circular scenario in Fig.~\ref{fig:wf_ecc0}. 
The residuals become insensitive to adding six overtones already at $\tau\sim \tau_{\rm LR}+5$ and soon after all the other overtones quickly decay out of the strain.
We argue this is a consequence of radial infalls being characterized by faster dynamics, and the absence of an oscillating, driving source, suppressing the excitation of high overtones.

\clearpage

\section{Activation and impulsive coefficients and contributions to the waveform}\label{app:coeffs}

\begin{figure*}
\centering
\includegraphics[width=\bigfigwidth]{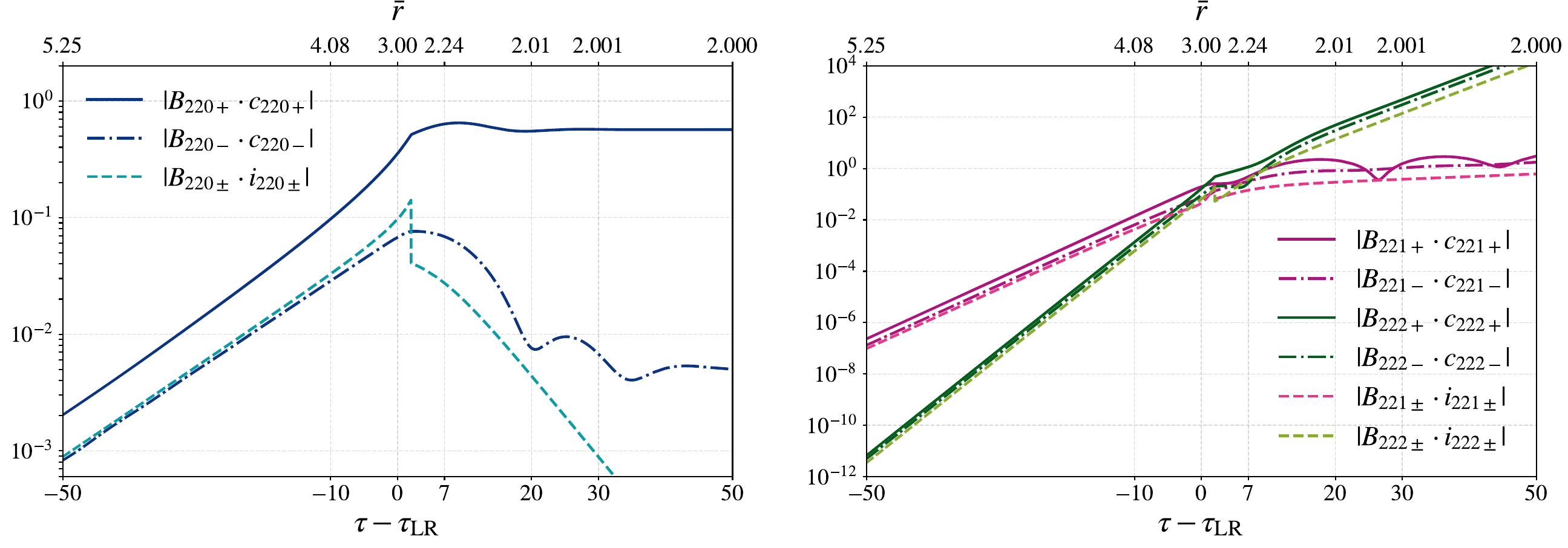}
\includegraphics[width=\bigfigwidth]{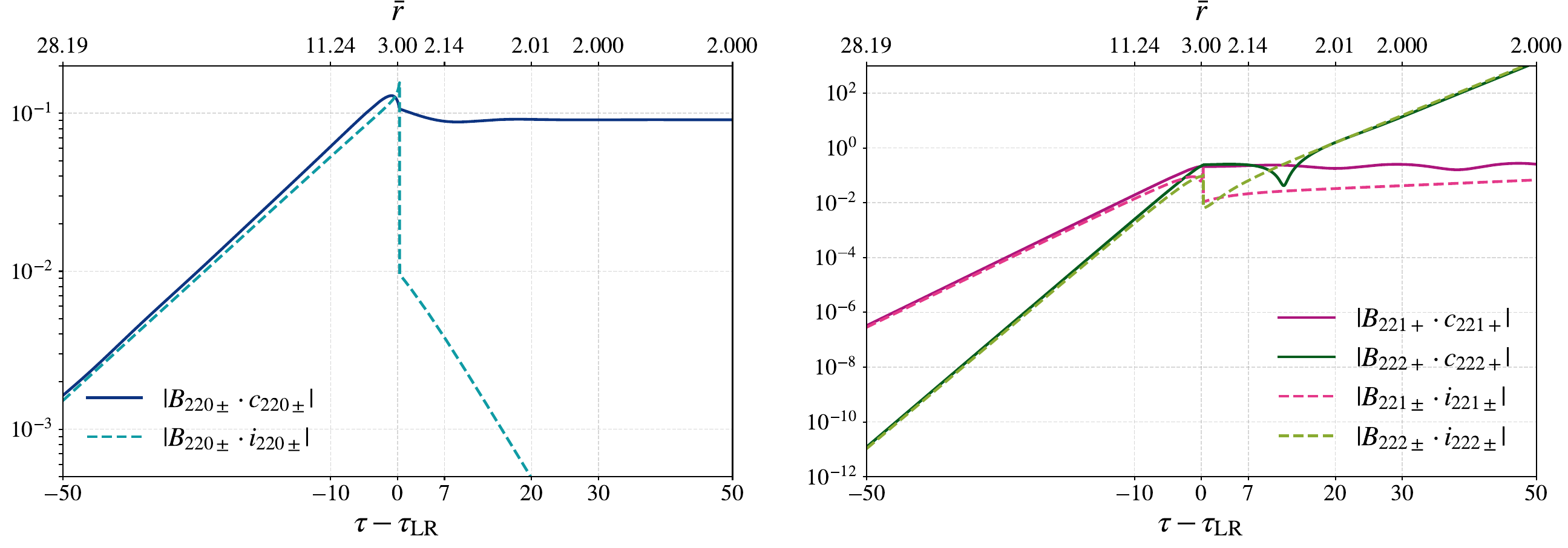}
\caption{
\textbf{Top}: Absolute value of the QNM activation $c_{22n\pm}$  
and impulsive $i_{22n\pm}$ coefficients
(weighted with the geometric excitation factors $B_{22n\pm}$) of the modes $220\pm$ (\textit{left}) and $221\pm,\,222\pm$ (\textit{right}), vs the retarded time of the observer $\tau$ from the apparent light-ring crossing, $\tau_{\rm LR}$, for a \textbf{quasi-circular inspiral}.
On the top horizontal axes we show the apparent location $\bar{r}$ of the particle, defined in Eq.~\eqref{eq:causality_trajectory}.
\textbf{Bottom}: Same as in top row, for a \textbf{radial infall} from $r_0=50$ with particle initial energy $E_0=1$.
}
\label{fig:c22n_i22n_vs_tau}
\end{figure*}

\begin{figure*}[t]
\centering
\includegraphics[width=0.82\bigfigwidth]{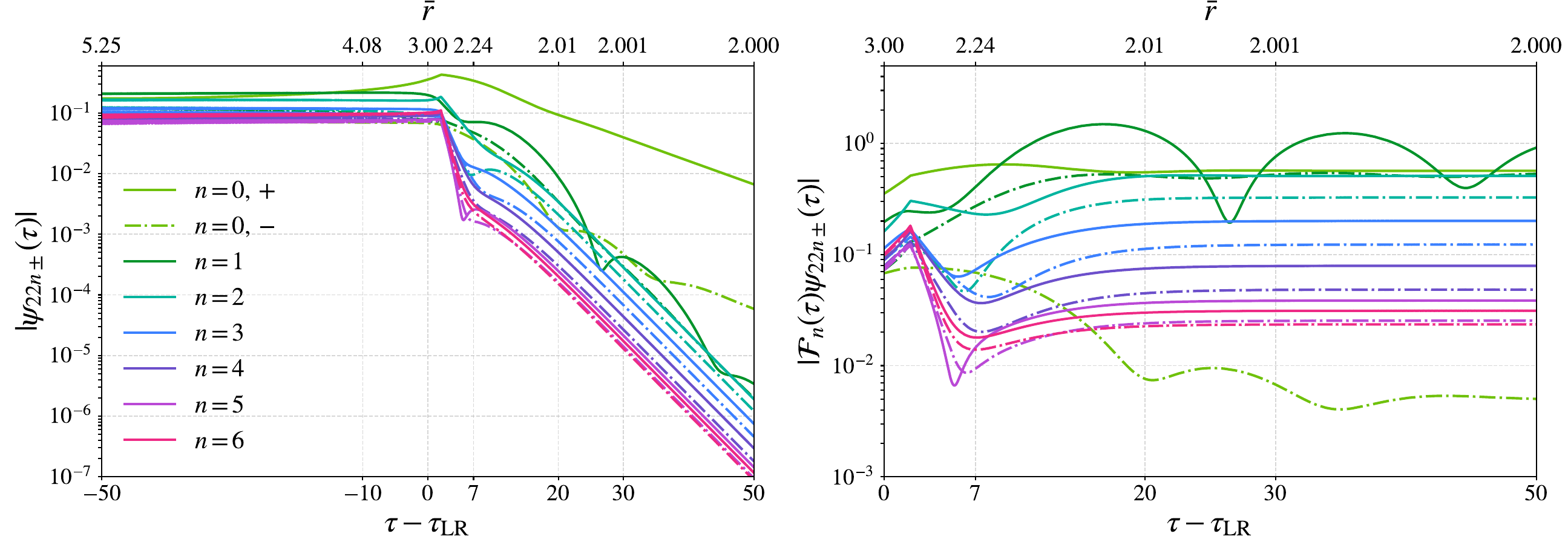}
\includegraphics[width=0.82\bigfigwidth]{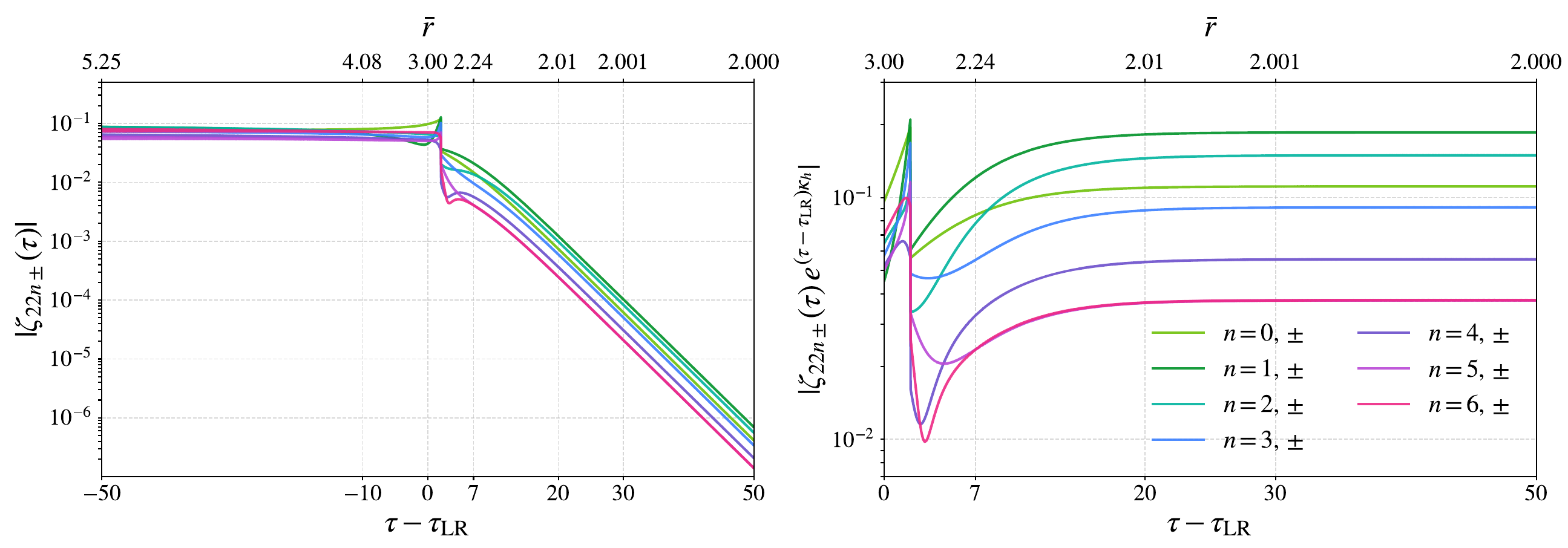}
\caption{
\textit{Left:} Activation (top) and impulsive (bottom) waveforms vs retarded time of the observer at $\mathcal{I}^+$ measured from the light-ring crossing time, for a \textbf{quasi-circular inspiral}; different colors refer to different overtone number $n$, while dashed lines are for mirror modes. 
\textit{Right:} Same quantities, rescaled with their late time behavior, see Eq.~\eqref{eq:func_rescaling}.}
\label{fig:psi_zeta_ecc0}
\end{figure*}

\begin{figure*}[t]
\centering
\includegraphics[width=0.82\bigfigwidth]{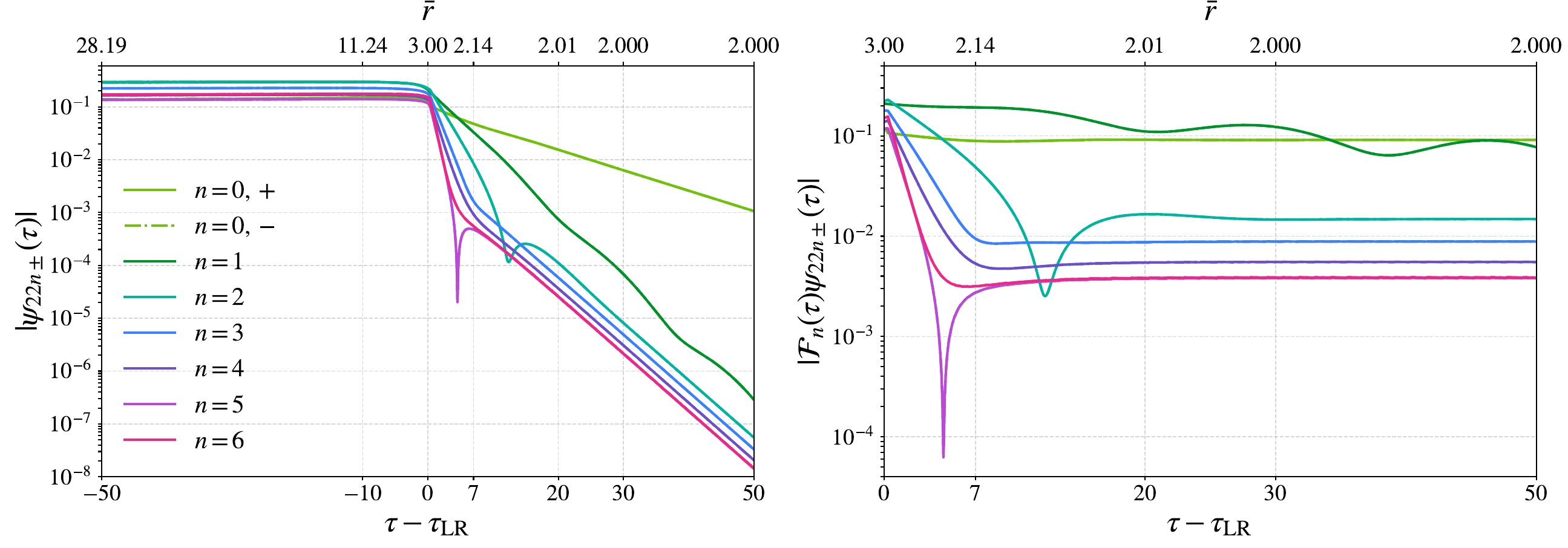}
\includegraphics[width=0.82\bigfigwidth]{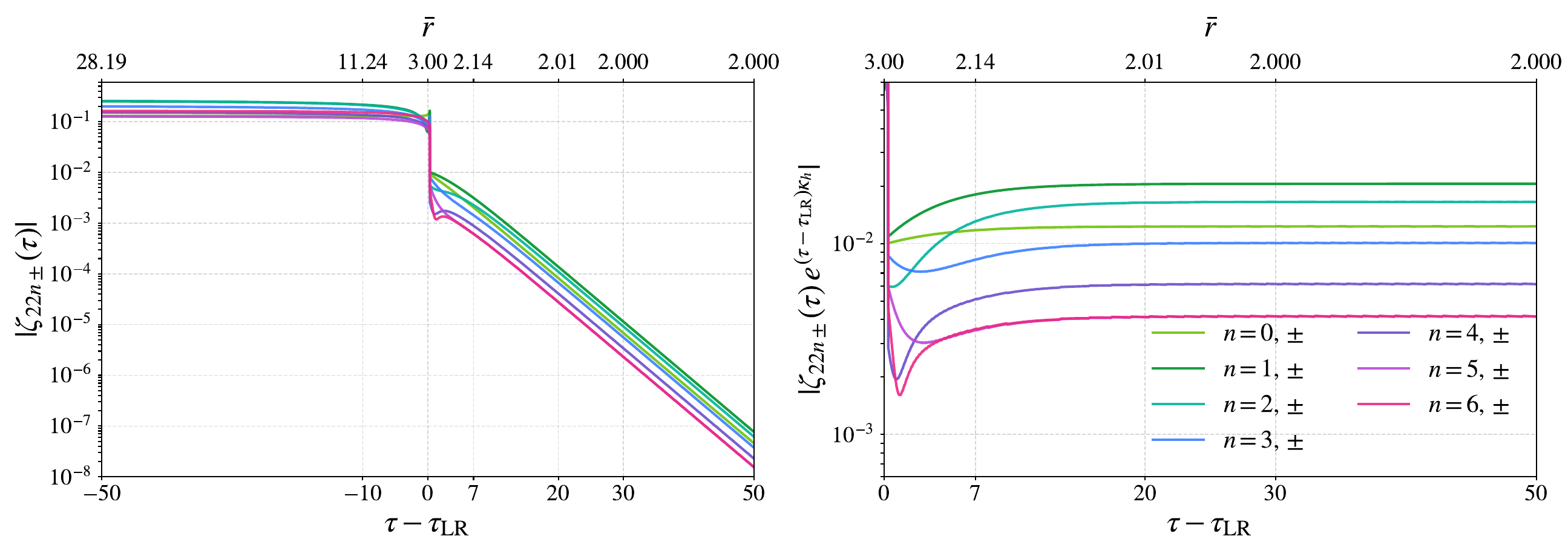}
\caption{
\textit{Left:} 
Same as in Fig.~\ref{fig:psi_zeta_ecc0} but for a \textbf{radial infall} from $r_0=50$, with initial specific energy $E_0=1$; different colors label overtone numbers $n$, dashed lines mirror modes. 
\textit{Right:} Same quantities, rescaled with their late time behavior, see Eq.~\ref{eq:func_rescaling}.}
\label{fig:psi_zeta_radinf}
\end{figure*}

The excitation $c_{22n\pm}$ and impulsive $i_{22n\pm}$ coefficients are shown in Fig.~\ref{fig:c22n_i22n_vs_tau},
for both the quasi-circular trajectory and radial infall.
In the latter, regular and mirror modes coincide in magnitude due to symmetry.
The results are very similar to those of Paper I; the coefficients grow exponentially at early times, and there is a change in behavior near the light-ring crossing.
If in Paper I the change was gradual, when using the condition in Eq.~\eqref{eq:new_caus} the change is sharp (not differentiable) at $\bar{r}_*=0$.
This was to be expected, since the bounce time in Paper I interpolates, while crossing the light ring, between the scattering and the light-cone condition.
Instead, the prescription used in this work switches sharply between the two at $r_*=0$.
The other significant difference is that now the impulsive coefficients decay to zero at a much faster rate. 
Finally, we briefly mention that for the overtones, the late-time behavior is still dictated by the redshift terms~\cite{DeAmicis:2025xuh,Rosato:2026moe}, which are suppressed for radial infalls due to a faster plunge dynamic.
Following Paper I, we define the activation $\psi_{n\pm}$ and impulsive $\zeta_{n\pm}$ contribution of each QNM to the full waveform as
\begin{equation}
    \psi_{n\pm}(\tau)\equiv B_{n
    \pm}\, c_{n\pm}(\tau)\,e^{-i\omega_{n\pm}(\tau-\rho_+)} \, ,
    \label{eq:psi_def}
\end{equation}
\begin{equation}
    \zeta_{n\pm}(\tau)\equiv B_{n
    \pm}\, i_{n\pm}(\tau)\,e^{-i\omega_{n\pm}(\tau-\rho_+)} \, ,
    \label{eq:zeta_def}
\end{equation}
In Figs.~\ref{fig:psi_zeta_ecc0} and~\ref{fig:psi_zeta_radinf}, we show their behavior as a function of the retarded time $\tau$ of the observer at $\mathcal{I}^+$.
We also show these quantities rescaled with their dominant late-time behavior~\cite{DeAmicis:2025xuh}
\begin{equation}\label{eq:func_rescaling}
\begin{split}
 &\mathcal{F}_{n=0}\equiv e^{|\mathrm{Im}(\omega_0)|\, \tau}\, ,\\
& \mathcal{F}_{n>0}\equiv e^{\tau/4}\, .
\end{split}
\end{equation}
The results are overall consistent with those in Paper I, with some differences.
When using the prescription in Eq.~\eqref{eq:new_caus}, the contribution of each mode is approximately constant in time before the light-ring crossing. At the crossing of $r_*=0$, corresponding to $r\simeq 2.557$, there is a change in behavior: the contribution of the fundamental mode (and its mirror mode) decays exponentially at a rate given by the imaginary part of the quasinormal frequency. 
Overall, compared to Paper I, the contributions evolve much more slowly before bounce-crossing, when they transition sharply.
For the quasi-circular trajectory (and other non-extreme eccentricities) the overtone contributions instead transition rapidly to a redshift decay.
In the radial infall case, the redshift is suppressed with respect to the QNM behavior so that the fundamental mode and its mirror mode are dominated by the quasinormal decay even at late times ($\tau\approx\tau_{\rm LR}+50$).
The main difference with respect to Paper I is that before the $\bar{r}_*=0$-crossing, waveform contributions do not exhibit a monotonic suppression with the overtone number $n$.
This results in a slow convergence in the QNM-propagated signal with the number of overtones before the $\bar{r}_*=0$ crossing.


\clearpage

\section{Single-mode-propagated waveform}\label{app:single_mode}

\begin{figure*}[t]
\centering
\includegraphics[width=0.95\textwidth, trim={0cm 0.9cm 0cm 0.7cm}]{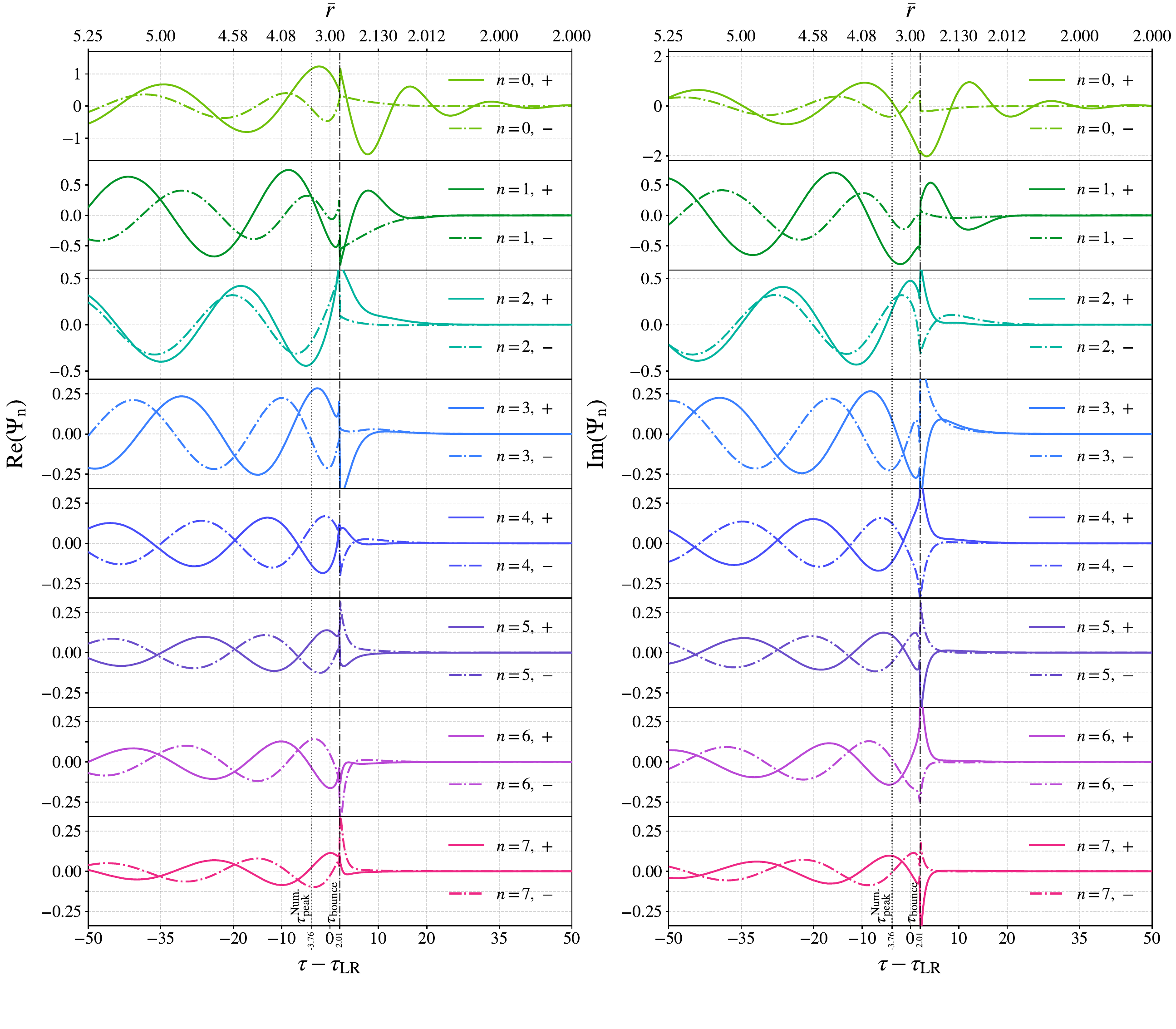}
\caption{QNM-propagated signal for each mode overtone $n$ vs the retarded time $\tau$, shifted with respect to the light ring crossing time $\tau_{\rm LR}$.
Results relative to a \textbf{quasi circular inspiral}.
}
\label{fig:overt_phases_re_im}
\end{figure*}

In Fig.~\ref{fig:overt_phases_re_im}, we show the real and imaginary part of the signal propagated by each QNM. As discussed in the main text around Fig.~\ref{fig:overt_phases}, the contribution of each mode during the inspiral-plunge (before the $\bar{r}_*=0$ crossing) is progressively more and more in counter phase as the overtone number increases. 

\clearpage

\bibliography{bibliography}

\end{document}